\documentclass[preprint,12pt]{elsarticle}
\pdfoutput=1

\usepackage[T1]{fontenc}

\usepackage[portuguese]{babel}

\usepackage{amssymb}

\usepackage{caption}
\usepackage{subcaption}
\usepackage{float} 

\begin{document}

\begin{frontmatter}



\title{Outdoor Systems Performance and Upgrade}


\author[inst1]{L. Lopes}
\author[inst2]{S. Andringa}
\author[inst2,inst3]{P. Assis}
\author[inst1]{A. Blanco}
\author[inst1]{N. Carolino}
\author[inst7]{M. A. Cerda}
\author[inst4]{F. Clemêncio}
\author[inst2,inst3]{R. Conceição}
\author[inst1]{O. Cunha}
\author[inst6]{C. Dobrigkeit}
\author[inst2]{M. Ferreira}
\author[inst5]{C. Loureiro}
\author[inst2]{L. Mendes}
\author[inst2]{J. C. Nogueira}
\author[inst1]{A. Pereira}
\author[inst2,inst3]{M. Pimenta}
\author[inst1]{J. Saraiva}
\author[inst2]{R. Sarmento}
\author[inst8]{P. Teixeira}
\author[inst2,inst3]{B. Tomé}

\affiliation[inst1]{organization={Laboratório de Instrumentação e Física Experimental de Partículas},
            addressline={Departamento de Física da Universidade de Coimbra},
            city={Coimbra},
            postcode={3004-516},
            state={Coimbra},
            country={Portugal}}

\affiliation[inst2]{organization={Laboratório de Instrumentação e Física Experimental de Partículas},
            addressline={Av. Professor Gama Pinto, n. 2, Complexo Interdisciplinar (3is)},
            city={Lisboa},
            postcode={1649-003},
            state={Lisboa},
            country={Portugal}}

\affiliation[inst3]{organization={Departamento de Física, Instituto Superior Técnico, Universidade de Lisboa},
            addressline={Avenida Rovisco Pais, n. 1},
            city={Lisboa},
            postcode={1049-001},
            state={Lisboa},
            country={Portugal}}

\affiliation[inst4]{organization={Escola Superior de Saúde do Politécnico do Porto},
            addressline={Rua Valente Perfeito, 322},
            city={Vila Nova de Gaia},
            postcode={4400-330},
            state={Porto},
            country={Portugal}}

\affiliation[inst5]{organization={LIBPhys},
            addressline={Department of Physics, University of Coimbra},
            city={Coimbra},
            postcode={3004-516},
            state={Coimbra},
            country={Portugal}}

\affiliation[inst6]{organization={Universidade Estadual de Campinas},
            addressline={IFGW},
            city={Campinas},
            state={São Paulo},
            country={Brasil}}

\affiliation[inst7]{organization={Observatótio Pierre Auger},
            city={Malargue},
            country={Argentina}}

\affiliation[inst8]{organization={Physics Department (ECT), Institute of Earth Sciences (ICT/IIFA), Earth Remote Sensing Laboratory (EaRSLab), University of Évora},
            addressline={Rua Romão Ramalho no. 59},
            city={Évora},
            postcode={7000-671},
            state={Évora},
            country={Argentina}}

\begin{abstract}
Over the last two decades, the possibility of using RPCs in outdoors systems has increased considerably. Our group has participated in this effort having installed several systems and continues to work on their optimization, while simultaneously studying and developing new approaches that can to use of RPCs in outdoor applications.

In particular, some detectors were deployed in the field at the Pierre Auger Observatory in 2019 remained inactive, awaiting the commissioning of support systems. During the pandemic the detectors were left without gas flow for more than two years, but were recently reactivated with no major problems.

The LouMu project combines particle physics and geophysics in order to map large geologic structures, using Muon Tomography. The development of the RPC system used and the data from the last two years will be presented.

Finally, recent advances in a large area (1 m$^2$) double gap-sealed RPC will be presented.

\end{abstract}



\begin{keyword}
Sealed RPCs \sep RPC Outdoor operation \sep Low gas consumption \sep Cosmic rays \sep Gaseous Detectors \sep Geophysics.
\PACS 0000 \sep 1111
\MSC 0000 \sep 1111
\end{keyword}

\end{frontmatter}


\section{Introduction}
\label{sec:sample1}
Resistive Plate Chambers (RPCs) \cite{Santonicoetal1981} can be found in a large number of experiments, mainly in indoor but also in outdoor \cite{AIELLI200692,AGNETTA199664,Ambrosino_2014,gi-2-55-2013}. The possibility of using RPCs outdoor generates large interest inside our group. In 2013  we started the development of the first prototypes \cite{L_Lopes_2013} and, since then, a large number of detectors have been assembled and deployed in different locations \cite{Lopes_2016,Lopes_2020,Garcia-Castro:2021vku,Alvarez-Pol_2015,Abreu:2017vsj}.

Our main line of research focuses on different aspects of the gas system, such as the reduction of gas consumption or its possible simplification. This effort is motivated by the need to cover large areas with dispersed stations in Antiparticle Experiments, which makes the use of a centralized gas system unviable, being the only solution a large number of small simple systems.  In this line, the use of a mono-component gas, tetrafluorethane, as a counting gas is of major importance. This might downgrade slightly the performance of the detector when compared to "standard mixture". But if the timing precision requirement could be higher than 100 ps, it has been proved that the stability and robustness are not affected \cite{Lopes_2016}, and this option is very attractive due to its simplicity and low cost.  The reduction of gas consumption is also of major interest, both from a financial, logistically and environmental point of view. In this respect, stable operation with a gas flow rate of less than 5 cc.min${^-1}$ for 2 m$^{^2}$ 1 mm double gap chambers (corresponding to 1 volume change per day) is an important achievement. Last but not least, the possibility to operate RPCs without gas flux is one of our main challenges.

In 2019, efforts to assemble a fully  sealed (zero gas flow) chamber were successful and a 2 mm single gaps RPC was operated stably for more than 6 months without any evidence of gap seal degradation \cite{Lopes_2020}. Unfortunately, the role of the ambient relative humidity (RH) in the total current consumed by the detector is clear. But due to pandemic, minor advances were done to solve this problem.

This communication describes progress on different projects concerning the outdoor operation of the RPC system, namely: the installation of an engineering array in the infill region of the Pierre Auger observatory, the performance of a RPC based muon telescope used  for mapping geological structures, and the development of large-area multigap RPCs.

\section{MARTA Engineering Array at Pierre Auger observatory}
\label{sec:sample2}
\begin{figure}[h!]
    \centering
    \includegraphics[scale=0.05]{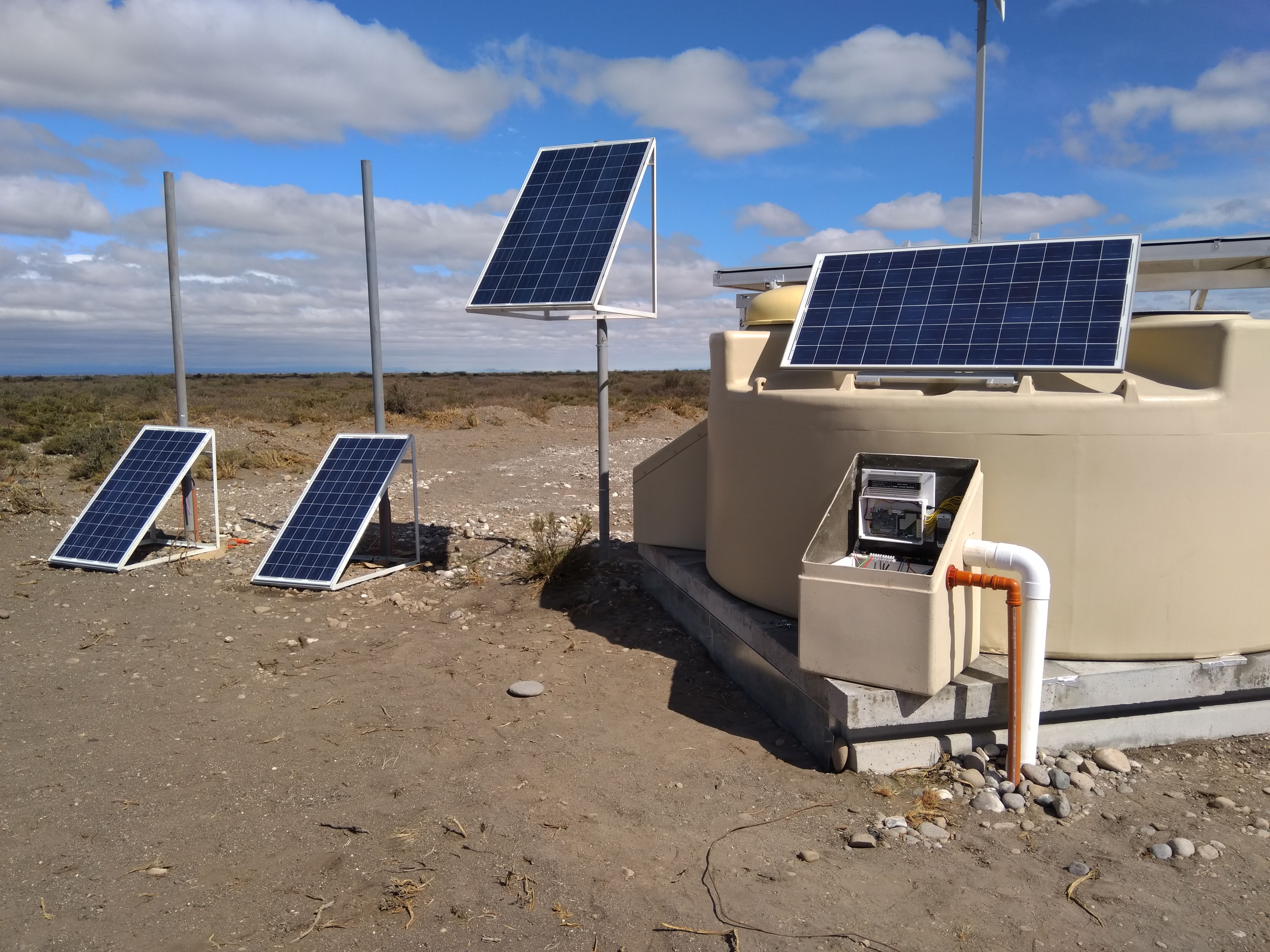}
    \caption{Peter Mazur station during deployment November 2019. The precast is underneath the tank and housing the four RPCs.}
    \label{fig:Peter Mazur}
\end{figure}

The Auger array is composed of Water Cherenkov Stations (WCS) at a distances 1500 m from their neighbors. In the central part of this array, the infill, which covers an area of 23.5 km2, the WCS are placed in a hexagonal grid at a distance of 750 m from their neighbors. The Muon Array with RPCs for Tagging Air showers (MARTA) project will upgrade one of the hexagons of the grid with a precast structure to support the WCS and four trigger RPCs underneath \cite{L_Lopes_2013,Lopes_2016}. In total seven of such structures will be deployed, six at the vertices together with one in the center forming the so call RPC Engineering Array. The deployment started at the end of 2019 with the upgrade of one of the WCS, the Peter Mazur station, \ref{fig:Peter Mazur}. The precast is located underneath the tank housing the four RPCs. The gas cylinder and the gas pipes were installed underground (not visible in the figure), to be less sensitive to extreme temperatures excursions during Summer and Winter.

The detectors were operated at a gas flow rate of 4 cc.min$^{-1}$, and a High Voltage (HV) working point of around 5100 kV.gap$^{-1}$ with operation current below 100 nA/RPC. Due to pandemic, once the gas cylinder was depleted, it was not possible to replace it and the detectors remained  for more than 2 years without gas flux or HV. In May 2022, It was possible to reactivate the installation. During the startup of the station, water was observed inside the gas pipes, which can only be explain by condensation. The gas pipes were cleaned and reconnected and after a few days of operation with fresh gas the HV of the RPCs was ramped up to nominal working point with less than 200 nA/RPC. The glues of the acrylic blubbers were also found to be deteriorated. Furthermore, the Polyamide pipes seems to suffer from the extreme temperature excursions and RH. To mitigate these problems metallic pipes will be installed and the acrylic blubber glues will be replaced by threaded parts.  Unfortunately, the lack of monitoring data limits our system knowledge, but fortunately after reactivation of the station the data acquisition  and the monitoring systems are now fully functional.

\section{Muon tomography in Lousal mine}
\label{sec:sample3}
Muon tomography in Lousal mine (LouMu) \cite{Teixeira_2022} is a project combining physics and geophysics groups with the aim of mapping large geological structures using the Muon Tomography technique. For this purpose, a four planes RPC telescope with an active are of 1 m$^2$ and an intrinsic efficiency close to 1 for muons, is used. One of the planes is readout with strips, while the other three are readout with 49 pads (38x38 mm$^2$) in the central area of 30x30 cm$^2$, used for muon tracking, and fifteen auxiliary pads around with different shapes used for testing purposes. Nevertheless, the whole active area is being monitored and RPCs practical quantities: HV, HV currents, charge, efficiency, streamer fractions and gas flow rate, and environment variables: atmospheric pressure, temperature and relative humidity (RH), are measured and recorded.

\begin{figure}[H]
    \centering
    \begin{subfigure}{0.45\textwidth}
    \includegraphics[scale=0.03]{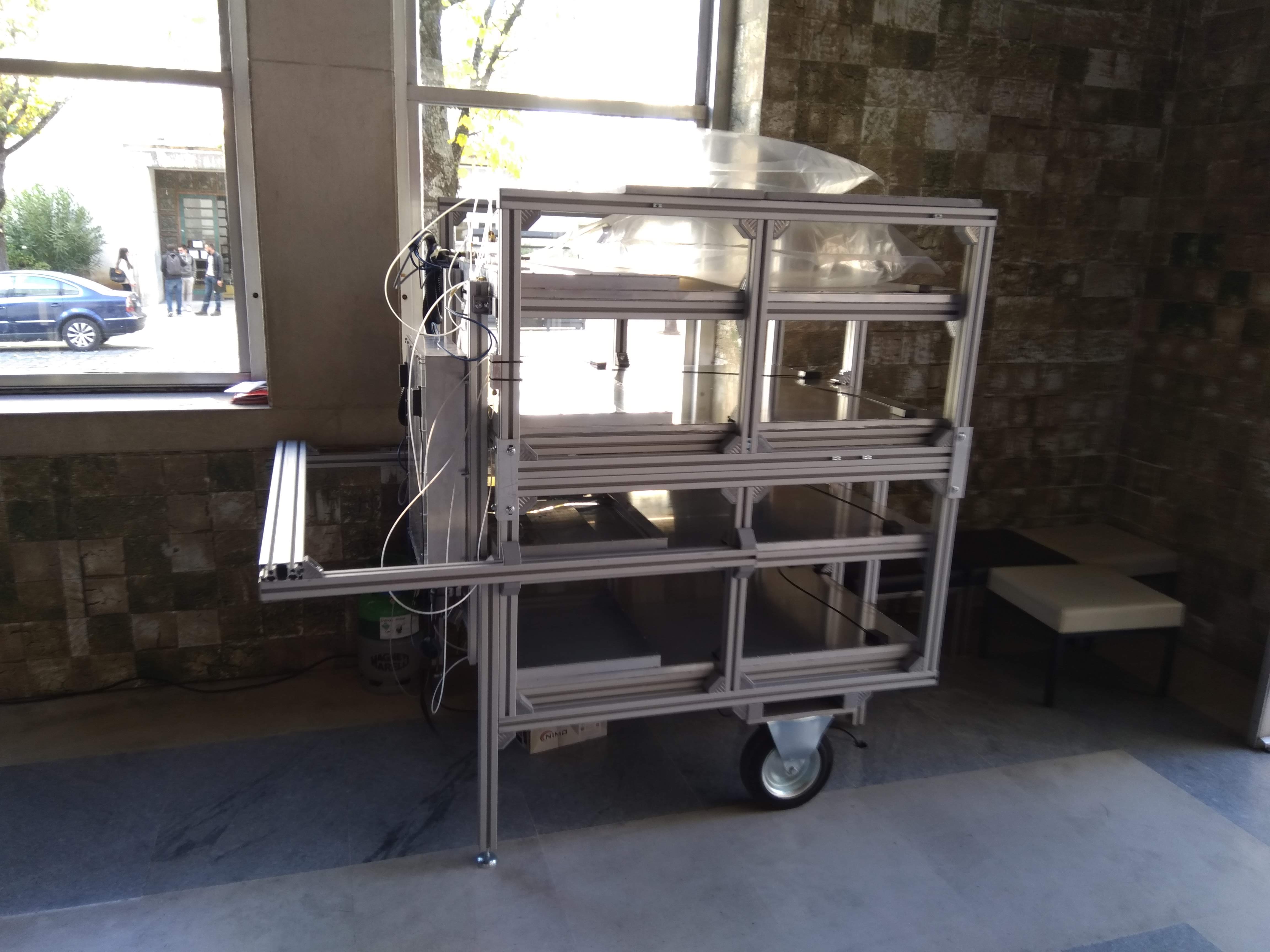}
    \caption{The hall of Physics Department}
    \label{fig:subim1}
    \end{subfigure}
    \begin{subfigure}{0.45\textwidth}
    \includegraphics[scale=0.03]{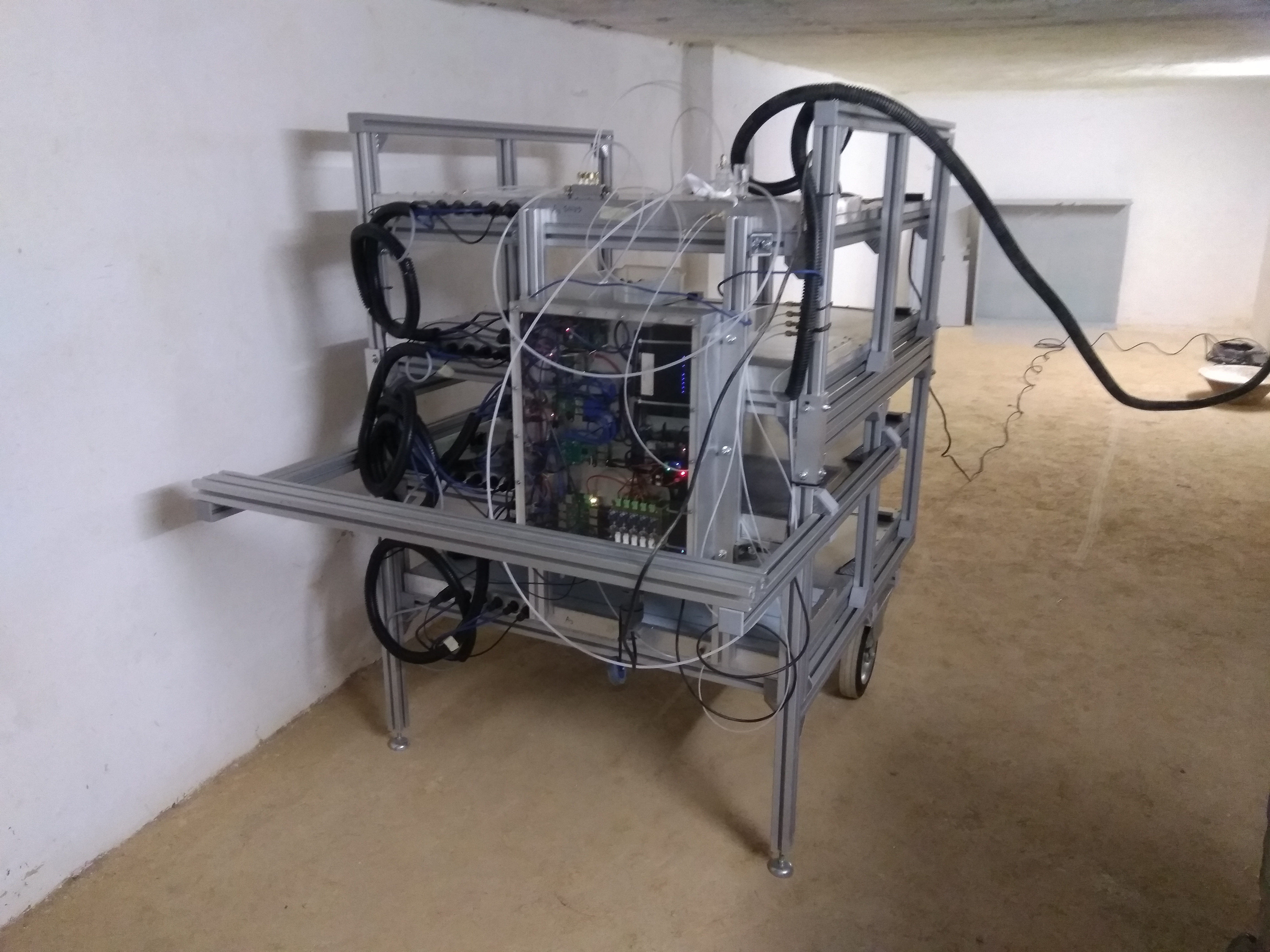}
    \caption{Gallery in Lousal mine}
    \label{fig:subim2}
    \end{subfigure}
    \caption{The telescope}
    \label{fig:image2}
\end{figure}

The telescope was assembled \ref{fig:subim1} at the end of 2019 and started data taking at the Coimbra Detector Laboratory in early 2020. The HV is automatically adjusted depending on temperature and absolute pressure variations to stabilize gain and, consequently, efficiency. The chambers are flushed with pure Tetrafluorethane at gas flow rates below 5 cc/min.plane. The gas at the outlet of each plane is collected in two Tedlar bags (total volume 200 L) and once the bags are filled, the gas is compressed inside a gas cylinder with an oil-free compressor for reuse or simply sent for recycling. This way, the gas losses are minimized and the system could be considered environmentally friendly, with little or no contribution to the global warming.

In May 2022 the telescope was moved to the Lousal mine \ref{fig:subim2}. Due to safety regulations, the gas cylinder had to stay out of the gallery, at a distance of more than 130 m from the telescope. For this reason, we were forced to use more that 140 m of polyamide pipe to supply the gas to the telescope and another 140 m to collect the gas to the Tedlar bags located close to the gas cylinder. After carefully flushing the pipes to remove all the air inside, they were connected to the telescope and the system was restarted.

\begin{figure}[H]
    \centering
    \begin{subfigure}{0.45\textwidth}
    \includegraphics[scale=0.25]{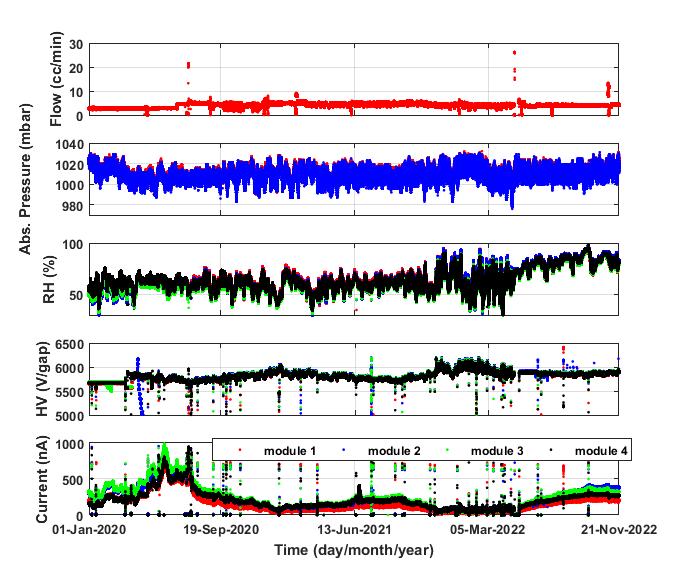}
    \caption{Environment monitoring}
    \label{fig:subim3}
    \end{subfigure}
    \begin{subfigure}{0.45\textwidth}
    \includegraphics[scale=0.25]{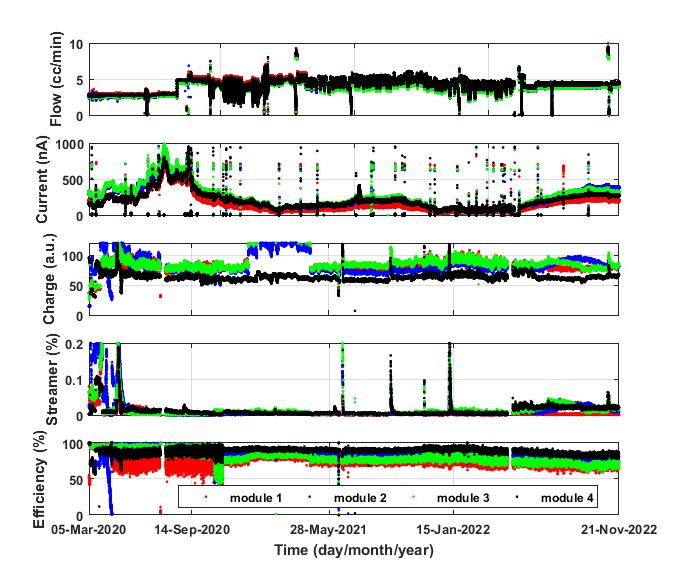}
    \caption{Chambers practical quantities}
    \label{fig:subim4}
    \end{subfigure}
    \caption{Environment variables and RPCs practical quantities}
    \label{fig:image3}
\end{figure}

After installation in early May, a continuous increase in the operation current, in all chambers, was observed, \ref{fig:image3}. This increase is correlated with a degradation of charge and streamer fraction together with a residual decrease on the efficiency, which is of major importance in this application. All these observations are valid for all four chambers, suggesting that the cause is a change in environmental conditions or in the quality of the gas that is injected in each of the planes. The gas flow rate is the same used in the laboratory, the temperature, very stable inside the gallery (around 18 ºC), and absolute pressure also are within normal values. Only the RH shows a correlation with HV current, although, in the laboratory operation higher RH values were observed without any correlation with the currents.

The main difference is the long (140 m) polyamide pipe used to supply the gas from the cylinder to the chambers. Polyamide pipes are cheap and frequently used in many RPCs or gas detector applications.  When using short pipe lengths, as in laboratory operation, or at high flow rates, the effect of the water and atmospheric gas permeability of polyamide is mitigated and the effect on chamber performance becomes negligible. In our case, the very low gas flow rate and long pipe, without any extra protection, enhance the permeability effect.
To confirm this theory, a test was carried out by placing the gas cylinder close to the telescope and recreating the laboratory conditions. Due to safety restrictions, the test could only run for two days. However,  the effect on the HV currents was evident, as can be seen at the end of the plots \ref{fig:image3}.. To definitively improve the situation, a PTFE (Teflon) pipe, with a much lower permeability to water and atmospheric gases ,will replace the polyamide pipe.

\section{Sealed Resistive Plate Chamber}
\label{sec:sample4}
From the two previous sections, it is clear that the gas quality plays an important role in the performance of the chambers, as is to be expected in any gaseous detector. One approach, to avoid external contamination, would be to keep the gas perfectly sealed and isolated inside the gaps, the sealed RPC concept. This is expected to be a solution for some applications with low particle flow. Other direct advantage is the drastic decrease in gas consumption with zero contribution to global warming, important factors in view of the increasing limitations in the production and use of HFCs and SF$_6$. It is important to mention the numerous efforts in the community to find possible substitutes for these gases. However, it is not clear that a usable option, both from a performance and cost point of view, is close to being found.

Three years ago the first sealed chamber was assembled and operated without problems for more than six months \cite{Lopes_2020}. The main issue with the chamber was leakage currents through the external surfaces, which were correlated with the RH around the chamber.

Recently, a double 1 mm gap chamber with an effective area of 1 m$^2$ has been built. Some changes were introduced compared to the first prototype, but retaining the main concept of this technology: keep gas only in contact with glass. The gaps are sealed in the same way but the HV is applied through acrylic paint on the outer glass surfaces and three layers of polyethylene were placed over the paint to confine the HV.

\begin{figure}[H]
    \centering
    \begin{subfigure}{0.45\textwidth}
    \includegraphics[scale=0.25]{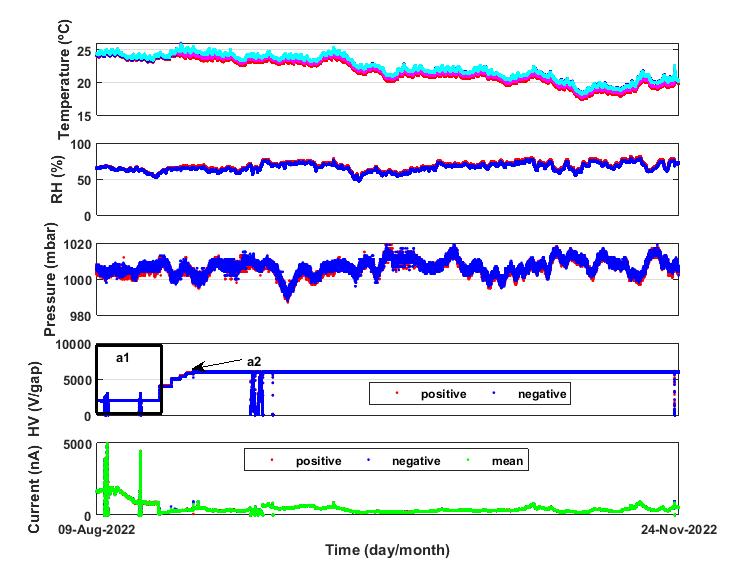}
    \caption{All monitored variables}
    \label{fig:subim5}
    \end{subfigure}
    \begin{subfigure}{0.45\textwidth}
    \includegraphics[scale=0.25]{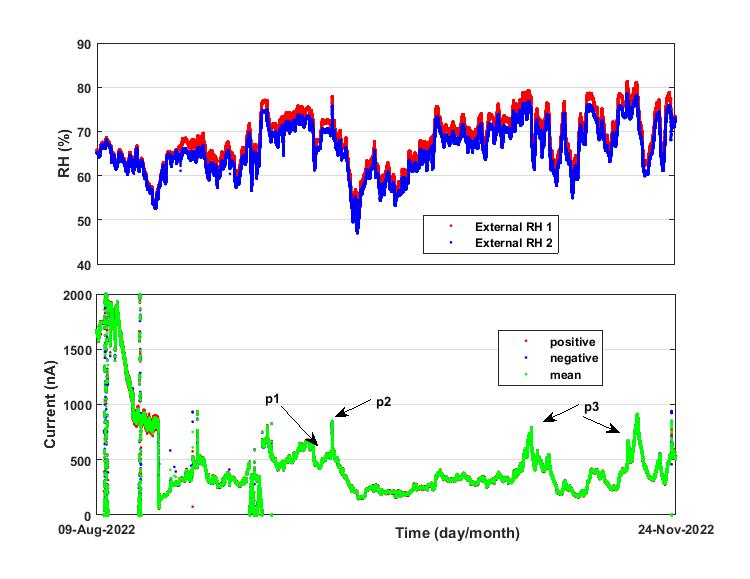}
    \caption{Dependence of the operation current of the relative humidity}
    \label{fig:subim6}
    \end{subfigure}
    \caption{Monitoring of the 1 m$^2$, double 1 mm gap-sealed Resistive Plate Chamber}
    \label{fig:image4}
\end{figure}

The HV current, HV, temperature, atmospheric pressure and RH are measured at a frequency of one minute, as shown in \ref{fig:subim5}. After assembly, the chamber was flushed with argon and a permanent discharge was establish as visible in \ref{fig:subim5} (small square labeled as a1). This procedure is performed to clean the gaps and evaluate their uniformity \cite{Lopes_2020}. After a couple of weeks, the gaps were filled with a mixture of 0.95(C$_2$H$_2$F$_4$) + 0.5(SF$_6$) and the HV slowly increased until the HV plateau was reached. The chamber was them sealed, \ref{fig:subim5} (label a2).

With only the HV current available as a practical quantity to evaluate the chamber status, it is not possible to be sure that the HV plateau has been reached. However, previous experience with similar chambers indicates that this is the correct HV value. As in the previous chamber, the HV leakage current at the external surfaces is considerable, and variations in HV current over time is correlated with the external RH, as shown in figure 4b, being much more visible for periods with low temperature excursions. To confirm this theory, an argon atmosphere was created around the chamber, \ref{fig:subim6} point labeled as p1, resulting in a fast decrease in the RH correlated with a similar decrease in the HV current. Another observable supporting this theory, is the sudden increase in HV current, \ref{fig:subim6} point labeled as p2, created by the presence of more than 30 people in the laboratory clearly correlated with a spike in the HR.  At the end of plot \ref{fig:subim6}, points labeled as p3, the increase in the HV current is caused by the switching of the room heating system. This is to be expected since the gain is not being adjusted as a function of temperature variations.

The quality of the gas inside the gaps was evaluated with a radioactive 60-Co source. Placing the source at the same distance create the same ionization current, 40 nA. Only small variations were observed due to the effect of temperature on the gain and probably also on the glass resistivity.

The readout electrodes has already been installed and complete characterization of the chamber will be done soon, measuring all practical quantities.

\section{Conclusions}
\label{sec:sample5}
he pandemic severely limited progress on the MARTA Engineering Array, however, it was possible to discover some situations that need to be improved, mainly related to the gas distribution system. In the coming months we expect to solve these problems and deploy some more stations.

The LouMu project underlines the importance of the material chosen for the gas distribution lines. Materials and choices that seemed suitable in the laboratory turned out to be completely wrong for outdoor applications. New PTFE pipes will be installed in the coming weeks to solve, or at least mitigate, the permeability problem detected in long polyamide pipes.

Sealed Resistive Plate Chambers could be an option, in certain low particle flow applications, to overcome the limitations imposed by the phase-out of the HFCs. Within the next year, it is expected that we could make considerable advances in the line of research.

\section*{Acknowledgement}
This work is supported by Portuguese national funds OE/FCT and by the Portuguese Republic; CERN/FIS-PAR/0012/2021, EXPL/FIS-OUT/1185/2021 and 2021/CERN/FIS/INS/0006.



 \bibliographystyle{elsarticle-num}
 \bibliography{cas-refs}





\end{document}